\documentclass[a4paper,twocolumn,11pt,accepted=2022-10-03]{quantumarticle}
\pdfoutput=1
\usepackage[utf8]{inputenc}
\usepackage[english]{babel}
\usepackage[T1]{fontenc}
\usepackage{amsmath}
\usepackage{hyperref}
\usepackage{tikz}
\usepackage{amsthm}
\usepackage{amsfonts}
\usepackage{comment}
\usepackage{xcolor}
\usepackage{amssymb}
\usepackage{graphicx}

\usepackage[caption=false]{subfig}

\theoremstyle{definition}

\theoremstyle{definition}
\newtheorem*{conjecture*}{Conjecture}

\begin{document}

\title{Design of quantum optical experiments with logic artificial intelligence}
\author{Alba Cervera-Lierta}
\affiliation{Chemical Physics Theory Group, Department of Chemistry, University of Toronto, Canada.}
\affiliation{Department of Computer Science, University of Toronto, Canada.}
\affiliation{Barcelona Supercomputing Center, Barcelona, Spain}
\orcid{0000-0002-8835-2910}
\email{alba.cervera@bsc.es}
\author{Mario Krenn} 
\affiliation{Chemical Physics Theory Group, Department of Chemistry, University of Toronto, Canada.}
\affiliation{Department of Computer Science, University of Toronto, Canada.}
\affiliation{Vector Institute for Artificial Intelligence, Toronto, Canada.}
\affiliation{Max Planck Institute for the Science of Light (MPL), Erlangen, Germany}
\orcid{0000-0003-1620-9207}
\email{mario.krenn@mpl.mpg.de}
\author{Al\'an Aspuru-Guzik}
\affiliation{Chemical Physics Theory Group, Department of Chemistry, University of Toronto, Canada.}
\affiliation{Department of Computer Science, University of Toronto, Canada.}
\affiliation{Vector Institute for Artificial Intelligence, Toronto, Canada.}
\affiliation{Canadian Institute for Advanced  Research  (CIFAR) Lebovic Fellow, Toronto,  Canada}
\orcid{0000-0002-8277-4434}
\email{alan@aspuru.com}
\maketitle

\begin{abstract}
Logic Artificial Intelligence (AI) is a subfield of AI where variables can take two defined arguments, True or False, and are arranged in clauses that follow the rules of formal logic. Several problems that span from physical systems to mathematical conjectures can be encoded into these clauses and solved by checking their satisfiability (SAT). 
In contrast to machine learning approaches where the results can be approximations or local minima, Logic AI delivers formal and mathematically exact solutions to those problems.
In this work, we propose the use of logic AI for the design of optical quantum experiments. We show how to map into a SAT problem the experimental preparation of an arbitrary quantum state and propose a logic-based algorithm, called \textsc{Klaus}, to find an interpretable representation of the photonic setup that generates it. We compare the performance of \textsc{Klaus} with the state-of-the-art algorithm for this purpose based on continuous optimization. We also combine both logic and numeric strategies to find that the use of logic AI improves significantly the resolution of this problem, paving the path to developing more formal-based approaches in the context of quantum physics experiments.
\end{abstract}

\section{Introduction}

The emergence of artificial intelligence (AI) has led to the proposal of alternative ways to tackle hard non-analytical problems. The AI canonical approach comes in the form of inductive generalizations through the use of big data, the well-known and established machine learning (ML) field. Although ML grounds rely on mathematical theorems related to continuous function representation, its probabilistic nature usually does not yield performance guarantees, even less understanding about why it works (or not) in a particular problem. Despite the progress in unraveling the learning paths of ML algorithms, ML sibling, logic AI \cite{mccarthy1960programs,nilsson1994probabilistic,darwiche2020three}, has the intrinsic potential of providing the validity and consistency of the answers we seek.

Logic AI is a subfield of AI that uses symbolic representation in the form of Boolean variables to extract formal deductions. In its basic form, it consists of encoding a set of rules into Boolean instances which validity can be checked with, for instance, satisfiability (SAT) solvers. 
The recent advances in SAT solvers have allowed the automatic resolution of extremely complex problems involving thousands of variables \cite{fichte2020time}. Long-standing conjectures such as the Boolean Pythagorean triples problem \cite{heule2016solving}, 
the Keller's conjecture (unresolved for 90 years) \cite{brakensiek2020resolution}, among others \cite{de2018chromatic,kaplan2021heesch,yolcu2021automated} have been solved using logic AI providing, in some cases, intricate, long \cite{lamb2016two} but correct deduction steps.  

In a quantum mechanical context, the use of logic AI has been slightly explored so far. A few examples propose a logic encoding and a SAT solver as an equivalent quantum circuit checker \cite{wille2013compact}, to find the mapping between a quantum circuit and a particular chip topology \cite{wille2019mapping} or to reduce the gate count \cite{meuli2018sat}. There are also works that find Boolean representations of quantum circuits \cite{ying2020symbolic}. These proposals use logic as a checker or optimizer. Here, we exploit logic AI for the design of quantum experiments. 

In this work, we propose a logic-based algorithm capable of designing a realistic quantum experiment. To be precise, our goal is to find a feasible photonic setup that generates an arbitrary quantum state. We benchmark our approach by comparing its performance with the best algorithm up to date, which is based on continuous numerical optimization, Theseus \cite{krenn2020conceptual}. To that aim, we will take advantage of the graph-theoretical representation that these setups can take, which can also be used for other quantum experiments such as gate-based quantum circuits, unitary operations generation or to design quantum error corrected photonic circuits \cite{bartolucci2021fusion}.

The structure of this paper is as follows. In the next section, we summarize the graph representation of optical setups and explain how to formulate a state preparation problem. In section \ref{sec:logic}, we show how to map the design problem into a set of propositional logic clauses. Section \ref{sec:Klaus} introduces the main algorithm, \textsc{Klaus}, that uses the logical instances presented in the previous section to find the minimal graph that corresponds to the optimal setup. In \ref{sec:benchmarks}, we benchmark \textsc{Klaus} and compare it with both the state-of-the-art algorithm Theseus and a hybrid algorithm proposal. Finally, we conclude and point to numerous exciting extensions of logical AI in quantum physics.

\section{Graph-based representation for quantum optics}\label{sec:PM_optical_setups}

\begin{figure*}[t!]
    \centering
    \includegraphics[width=\linewidth]{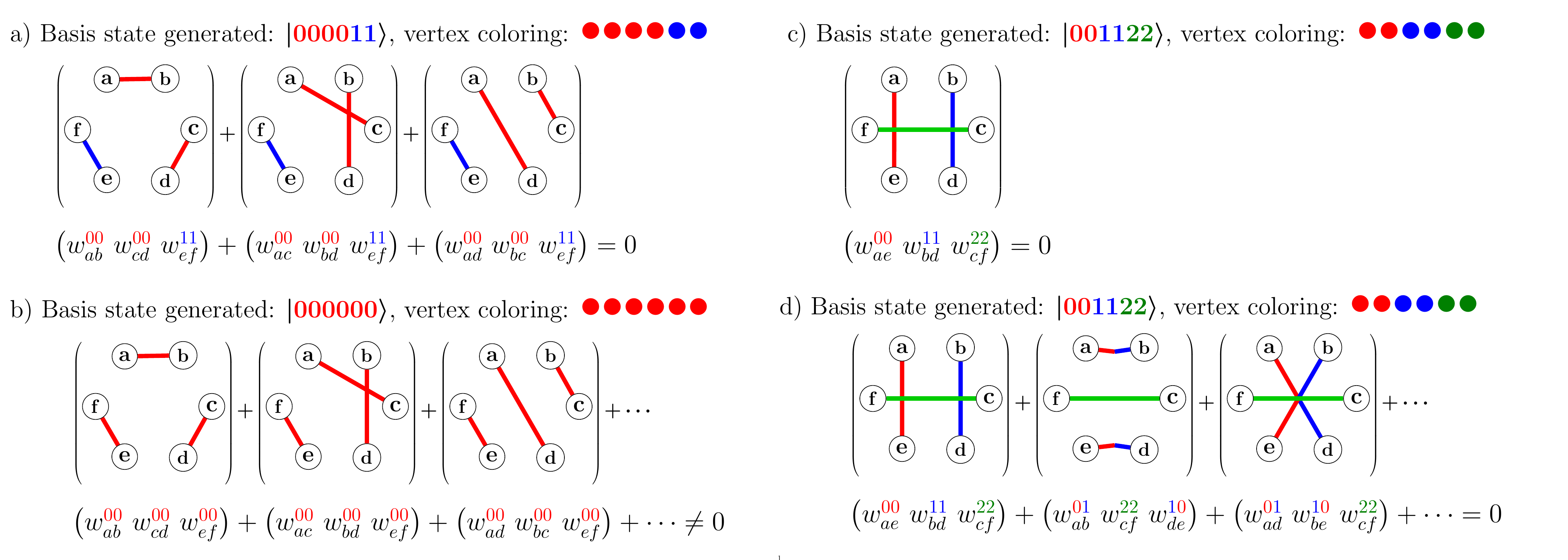}
    \caption{PMs equations examples for the generation of the $GHZ(6,3)$ state. \textit{a)} Two-colored PMs like the ones that generate the basis state $|000011\rangle$ can be canceled with each other by adjusting the weights of the edges. There are three PMs for each two-colored combination assuming monocolored edges. \textit{b)} To generate the basis state $|000000\rangle$ one needs to obtain a non-zero solution of the equation that sums the 15 PMs that generate that basis element. \textit{c)} For monocolored edges, the three-colored PMs are unique, which imposes that they must be zero. This imposes a very strong constraint that we will exploit later on for the logic encoding. \textit{d)} However, if we assume bicolored edges, there are 15 PMs for each color combination, including the three-colored ones. Thus it reduces the strength of the previous constraint.
    From these graphs, we can construct the photonic experimental setup following the mapping proposed in Ref.\cite{krenn2017entanglement}. For instance, each edge corresponds to an SPDC that generates a photon pair with the corresponding mode (color) in each path (letter).
    }
    \label{fig:obstruction}
\end{figure*}

A few years ago, a previously hidden bridge between quantum optical experiments and graph theory was discovered \cite{krenn2017quantum,gu2019quantumgraphsII,gu2019quantumgraphsIII} and has since been generalized as a highly efficient automated design algorithm for new quantum experiments \cite{krenn2020computer}. The underlying principle is that every quantum experiment can be described by an edge-colored weighted graph, and every graph stands for a quantum optical setup. In particular, every vertex of these graphs stands for a photon path (or a detector), every edge stands for a correlated photon path, the color represents the mode number and the complex edge weight stands for the amplitude of the photon pair. Such graphs can represent quantum states generated and transformed using linear optics, non-linear pair creation crystals, heralding and auxiliary photons, single-photon sources, photon number (non-)resolving detectors and others. 

The quantum state emerging from the experimental setup can directly be computed from the properties of the graph. A very commonly used technique in quantum optics conditions is the experimental result on the simultaneous detection of exactly one photon in each of the 
detectors \cite{pan2012multiphoton}. In the graph, this situation corresponds to a subset of edges that contain every vertex exactly once. This property of a graph is called a \textit{perfect matching} (PM). The final quantum state under this condition is then a coherent superposition of all PMs in the graph. A more detailed analysis of the equivalence between graph PMs and quantum states is presented in App. \ref{app:graphtostates}.

Given one of these graph representations, it will contain one or more PMs, each of them composed of different subsets of edges of different colors. As stated above, each of these edges represents a photon pair creation in the path (represented by the vertices) that they join. Each of these photons will have a mode represented by the color of the edge. This leads to the \textit{inherited vertex coloring} of the PM, i.e. we assign a color to each vertex corresponding to the color of the incident edge. The vertex coloring determines the basis element created in superposition with the other PMs vertex colorings. The amplitude of the basis element is determined by the \textit{weight of the PM}, i.e. the product of the weights of the edges. Different PMs can lead to the same vertex coloring but not necessarily the same PM weight. Thus, to compute the total amplitude of the basis element generated, one needs to sum all PM weights that generate that element, i.e. to compute the \emph{weight of the vertex coloring}. Since these weights can take complex values, they can cancel each other, thus having a set of PM with a given vertex coloring does not directly imply that the corresponding basis element is generated, as this interference may occur. 

Let's illustrate how can we set the quantum state preparation problem using an example of these graph representations. The formal definitions of this problem are provided in App. \ref{app:formalgraph}.

To generate a particular state, the weights of each vertex coloring that correspond to the basis states must match the state amplitudes, and the rest of the vertex coloring weights must be zero. Imagine that our goal is to generate the GHZ state of $n=6$ parties and $d=3$ dimensions, i.e., there are three possible different colors available (the 0, 1, and 2 modes). This state has three basis elements, each with amplitude $1/\sqrt{3}$:
\begin{equation}
    |\mathrm{GHZ}_{6,3}\rangle = \frac{1}{\sqrt{3}}\left(|000000\rangle + |111111\rangle + |222222\rangle\right).
\end{equation}
The general goal is the following: at least one of the contributions for the three basis elements must exist in the graph, while all other terms should vanish. 
Fig.\ref{fig:obstruction}a shows an example of a cancellation that must take place to cancel the generation of the basis state $|000011\rangle$, not present in the GHZ state. Fig.\ref{fig:obstruction}b, on the other hand, shows that the combination of PM with a unique coloring (in the figure, red) must be different from zero, in particular, it should be $1/\sqrt{3}$.
Notice that if we only assume monochromatic edges, there is only one PM for each tri-colored vertex coloring and, thus, the only possible solution is forcing this PM to be zero (Fig.\ref{fig:obstruction}c). However, if we allow bi-chromatic edges, there could be more tri-colored PM, allowing cancellations as in the bi-colored cases (Fig.\ref{fig:obstruction}d).

A mathematical conjecture has been proposed that states physically that it is not possible to generate a high-dimensional GHZ state with 6 or more photons with perfect quality and finite count rates without additional resources (such as auxiliary photons). Mathematically, this is equivalent to the question of whether there exists a weighted graph with at least three different vertex colorings of one color each \cite{DustinMixonGraph,krenn2019questions}, e.g. for $n=6$ a graph with PMs with all paths either blue, green or red and any other vertex coloring canceled out. The special case for positive weights was solved in 2017 by Ilya Bogdanov \cite{Bogdanov, krenn2017quantum}, but the case for negative and complex weights is open and contains the exciting possibility of using intricate quantum interference effects as a resource in quantum state preparation and transformation in quantum optics.
The question can be translated into a set of $d^n$ coupled nonlinear equations with $\frac{n(n-1)}{2}d^2$ complex variables \cite{MathoverflowEquationSystem}. The algebraic question is whether there exist solutions to this equation system for $n\geq6$ and $d\geq3$ and complex finite weights. The conjecture reduces to the simple statement that the equation system has no solution.

The emergence of obstructions such as the one shown in Fig.\ref{fig:obstruction}c suggests that combinatorics may play an important role in the generation of quantum states using this methodology. It is precisely the combinatorial nature of this problem that we will exploit with the help of a logic-based algorithm.

\section{Logic and SAT}\label{sec:logic}

In a Boolean algebra, the variables, called \textit{literals}, can take two Boolean values: True-False or 0-1. The available operations on these literals are \textit{disjunctions} $\lor$ (OR), \textit{conjunctions} $\land$ (AND) and \textit{negations} $\bar{x}$ ($\lnot$ or NOT).
Given a Boolean formula, the satisfiability (SAT) problem consist of finding a literals assignment that satisfies it, i.e. outputs True or 1. 

In the following subsections, we will encode the state preparation problem described in the previous section into a set of Boolean expressions whose satisfiability will give us a solution to the problem.

\subsection{Logic encoding}

\begin{figure*}[t!]
    \centering
    \includegraphics[width=\linewidth]{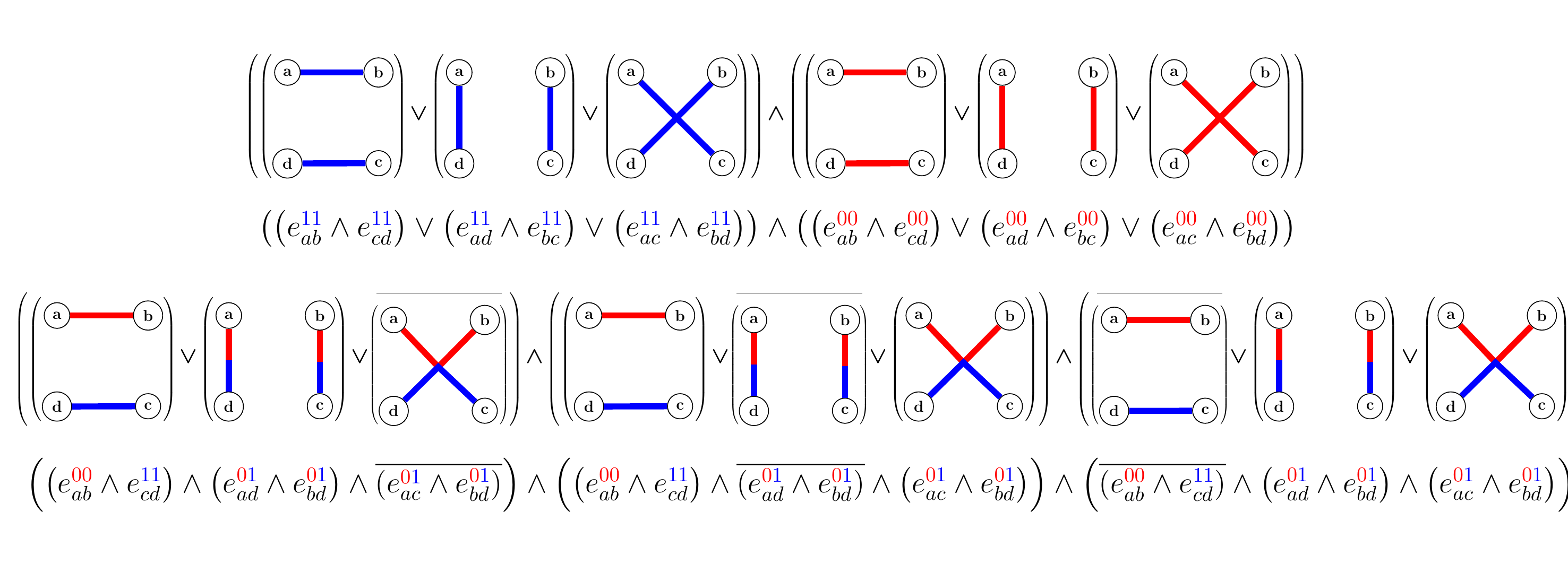}
    \caption{Logic example for the $GHZ$ state of $n=4$ and $d=2$. Assuming the edges can be bicolored, there are three possible PMs for each basis element. The Boolean variables are the edges of the graph $e_{ij}^{\alpha\beta}$ where $i,j$ correspond to the vertices and they are False if there is no edge and True otherwise. These weights also carry another degree of freedom, the color, which has as many dimensions $d$ as the state. The bar on top of a Boolean variable or expression corresponds to the negation of its value. Each PM is composed by the conjunction ($\land$) of all edges that compose it, so all edges must evaluate to True to have that PM. For those basis elements that appear in the target state, the logic instance corresponds to the disjunction ($\lor$) of all PMs; to evaluate to True, at least one of the PMs must exist, i.e. evaluate to True. This logic is represented in the top part of the figure, where the total expression must evaluate to True to obtain the superposition $|0000\rangle + |1111\rangle$. For those basis elements that are not in the target state, we can construct some obstructions. If all PMs except one evaluates to False, the remaining one has to be False as well. Other cases, like only one of them being False, can allow interference between the True PMs, a property not encoded in the logic. In the example (bottom part of the figure), the state $|0011\rangle$ must not appear, so the total expression must evaluate to True, as its negation will be added to the total set of clauses to be evaluated by the SAT solver.}
    \label{fig:logic_example}
\end{figure*}

We will explore the combinatorial nature of this graph problem to construct a set of logical clauses that can deliver a definite solution. 

In this problem, the literals will be each of the edges of the graph $e_{ij}^{\alpha\beta}$, where $(i,j)$ are the vertices joined by the edge (with $i<j$) and $(\alpha,\beta)$ are the inherited modes of the vertices, respectively. They will take the value True if they are present and False if they do not. Notice that we do not take into account that each edge can have a complex weight and thus there can be cancellations between PMs with the same vertex coloring.
Even though we do not encode the entire information and possibilities of the graph, we still get highly complex and powerful obstructions that we can use constructively in conjunction with SAT solvers.
This is by no means a restriction of representation. Negative and complex numbers can be represented by boolean variables effortlessly. As a simple example, we introduce another bit representing the sign of the number $s$ and the value bit $v$, such that numbers $-1,0,1$ for $11_b,00_b,01$. All boolean operations can be adjusted accordingly. Of course, in this way, we can also introduce more complex number systems such as fractions or complex numbers, but this is out of the scope of the current manuscript.

The logic clause to define a graph PM consist of replacing the PM weights by their corresponding (Boolean) edges and the products of the weights by $\land$. If one of the edges is False (there is no edge), the clause is False, and, therefore, we do not have that PM. The formal derivation of these clauses is presented in App.\ref{app:formallogic}. In the following paragraphs, we will show how this logic works using examples.

Let's start with a four-vertex graph example with six edges with the same color (mode 0) $\{e^{00}_{ab},e^{00}_{cd},e^{00}_{ac},e^{00}_{bd}, e^{00}_{ad},e^{00}_{bc}\}$. The logic clause that states the existence of the three PM $P_{1}$, $P_{2}$ and $P_{3}$ is
\begin{eqnarray}
    b_{P_{1}}(0,0,0,0)&=&e^{00}_{ab}\land e^{00}_{cd}, \nonumber\\
    b_{P_{2}}(0,0,0,0)&=&e^{00}_{ac}\land e^{00}_{bd}, \nonumber\\
    b_{P_{3}}(0,0,0,0)&=&e^{00}_{ad}\land e^{00}_{bc},
    \label{eq:exPM}
\end{eqnarray}
where $(0,0,0,0)$ represents the inherited vertex coloring (all photons are in mode 0). If only one of the edges in these PMs is False, that PM does not exist.

We require that at least one PM exist for each vertex coloring that appears in the target state. Following the previous example, if the state $|0000\rangle$ appears in the target state, then at least one of the previous PMs must be True:
\begin{multline}
    B(0,0,0,0)=b_{P_{1}}(0,0,0,0)\lor b_{P_{2}}(0,0,0,0) \\ \lor b_{P_{3}}(0,0,0,0).
\end{multline}
The above clause evaluates to True if at least one of the subclauses $b$ is True. 

If there are other basis elements in the target state, then all clauses of the form of $B$ must evaluate to True.
An example is shown in the top part of Fig.\ref{fig:logic_example}. The target state is the GHZ state of $n=4$ and $d=2$. There are two vertex colorings in the target state, the one corresponding to the $|0000\rangle$ basis element and the $|1111\rangle$ element, where the $|0\rangle$ and $|1\rangle$ states are represented in red and blue colors, respectively. Each PM is composed by two edges and, assuming the full-connected graph, there are three possible combinations: $\{(a,b),(c,d)\}$, $\{(a,c),(b,d)\}$ and $\{(a,d),(b,c)\}$. Since we want to generate a monocolored basis state, all edges have the same color on both ends. To obtain the two basis elements, at least one of the blue PMs and one of the red PMs have to evaluate to True. This is represented with the clause
\begin{equation}
    S = B(0,0,0,0) \land B(1,1,1,1).
\end{equation}

The remaining vertex colorings that do not appear in the target state must be False. However, as we mentioned before, the existence of more than one PM with a given coloring might be possible since there could induce a cancellation between the weighted PMs. The logic encoding that we propose cannot encode these cancellations, but we can include extreme cases independent of the weight values. We can have all PMs of a particular coloring and still obtain a cancellation between them, but if all PMs except one do not exist (they are False), the remaining one cannot exist either (should be False as well) because it cannot be canceled with anyone else. 

Figure \ref{fig:logic_example} bottom shows the clause for those PMs that generate the basis element $|0011\rangle$, which does not appear in the GHZ state. 

Let's analyze it piece by piece. The first part of the clause reads
\begin{equation}
    b_{P_{1}}(0,0,1,1)\lor b_{P_{2}}(0,0,1,1) \lor \overline{b_{P_{3}}(0,0,1,1)}. 
\end{equation}
If the three PMs exist (are True), this expression is True. If only two of them are True, the expression is still True. These two cases illustrate the fact that there could be cancellations between the PMs, so keeping them can be a solution once we search for the weights. If the first two PMs are False, the third one has to be False as well in order to keep the expression True. We must add the other two possibilities, i.e. that the other pairs of PMs are False, so the remaining one is forced to be False as well. This is why, the total clause shown in Fig. \ref{fig:logic_example} bottom contains three subclauses, to account for the permutations of PMs.

Altogether, the global set of clauses that encode the possible solutions for the generation of a particular state using graph PMs is a conjunction of clauses of type $S$, the ones that guarantee the existence of at least one PM for each target state basis element, and clauses of type $C$, a set of constraints on the PMs that should not appear in the final graph:
\begin{equation}
    K = S \land C.
    \label{eq:Klause}
\end{equation}
Given a set of edges, if $K=$False we can conclude it is not possible to obtain the target state. However, $K=$True does not guarantee the generation of this state due to the possible interference between PMs is not encoded in the clauses. For this reason, solutions such as the complete graph (all possible edges are True) output $K=$True, although heuristic optimization algorithms such as Theseus \cite{krenn2020conceptual} show that some states are not representable by graphs. For this reason, we mix these optimization strategies with \textsc{Klaus} to obtain and guarantee physical and interpretable solutions.

\begin{figure*}[th!]
    \centering
    \includegraphics[width=\linewidth]{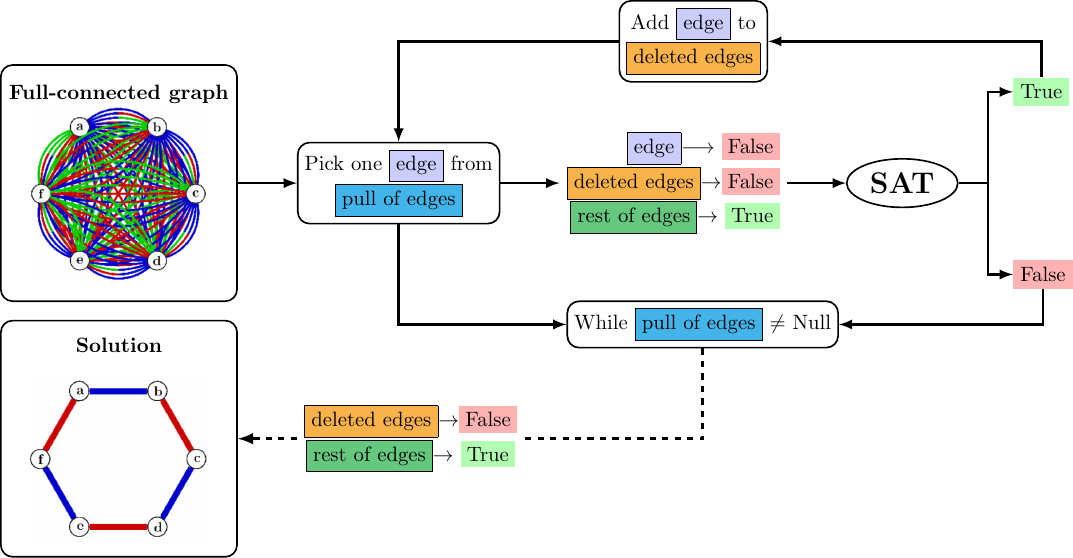}
    \caption{Diagram of \textsc{Klaus} algorithm. It starts with the complete graph, i.e., all edges are True. It randomly picks one of the edges, sets it to False, and checks if the set of clauses that encode the generation of the target state (see Eq.\eqref{eq:Klause}) is satisfiable using a SAT solver. If the SAT solver outputs True, the edge selected is apparently not required to generate the target state, so we can delete it, i.e., set it to False permanently. On the other side, if SAT is False, it means that an edge is required to generate the state, so it has to be True. The algorithm repeats the process of picking the other edges until all of them are classified as False or  True. As a result, we obtain a significantly sparsed graph. The final step consists of obtaining the graph weights that generate the required amplitudes to obtain the target state. This is done by numerically minimizing the infidelity of the graph obtained when replacing the edges with their corresponding weights.}
    \label{fig:klaus}
\end{figure*}

\subsection{Monochromatic edge obstructions}

The logical clauses presented in the previous section are general for both monochromatic and bi-chromatic edges. However, for graphs with only monochromatic edges, the problem simplifies substantially as the number of possible vertex coloring is much more constrained, therefore the logical approach is more powerful. One example is the one shown in Fig.\ref{fig:obstruction}c, where for the case of $n=6$ vertices tricolored vertex colorings are formed with unique PM. The same argument extends to more than three colors. In general, for a graph of $n$ vertices and monochromatic edges, vertex colorings composed by $d=n/2$ colors are unique. This fact implies that the condition of these vertex colorings is composed of a single clause: either that PM is True (if that coloring appears in the target state) or is False (if it does not). In the first case, it fixes the ``trueness" of all edges that form that PM. In the second case, it imposes that at least one of the edges must be zero. In either case, it could trigger a chain reaction on the rest of the clauses.

We test this approach to check if there exists a graph with monochromatic edges that generate the GHZ state of $n>4$ parties and $d\geq n/2$ local dimensions. We check if the set of clauses $K$ from Eq.\eqref{eq:Klause} is satisfiable, i.e. if there exists a solution for the literals that evaluates to True. We use the SAT solver from Mathematica language (which corresponds to MiniSAT in Mathematica 11). We obtained $K=$False for $n$ up to 8 and $d=n/2$ colors. For bigger systems, the amount of RAM required was out of range for our current computational capabilities. With these results we formulate the following conjecture to be added to other graph edge coloring conjectures such as the ones presented in Ref. \cite{krenn2019questions}:

\begin{conjecture*}
It is not possible to generate a graph $G$ with $n>4$ vertices and monochromatic edges each with one of $d\geq n/2$ possible colors, such that it contains single-colored PMs for each of these $d$ colors while no PMs with other vertex colorings are generated (or the amount of these PMs does not allow cancellations).
\end{conjecture*}

In the language of quantum state generation with photonic setups: it is not possible to generate exactly a GHZ state of $n>4$ parties and $d\geq n/2$ dimensions (and $n=4$ and $d>3$) using this graph approach without additional quantum resources (such as auxiliary photons).

\section{\textsc{Klaus} algorithm}\label{sec:Klaus}

SAT solvers look for a solution that is satisfiable no matter the number of True literals that it includes (looking for a particular solution will take exponential time), thus some of the solutions obtained may be cumbersome to interpret by humans. For instance, high-dense graphs with many True edges are allowed solutions of $K$ making it difficult to map them into a physical setup or to interpret the result to gain some understanding of how these states are physically generated. Moreover, the logic clauses do not provide the weights of the graph that generate the correct state amplitudes, so we need at least one extra step in our algorithm to compute these weights.
We propose a heuristic algorithm based on propositional logic named \textsc{Klaus} that aims to find a simplification of the satisfiable solutions of the logical clauses $K$ and to find the state amplitudes of the generated state.

Figure \ref{fig:klaus} shows the schematic representation of the \textsc{Klaus} algorithm. It starts with the fully connected graph, randomly selects one edge, and sets it to False. Then, it checks if $K$ is satisfiable using a SAT solver. If $K=$True, it means that edge was unnecessary to achieve the target state, so it ``deletes" it, i.e., sets it to False permanently. If $K=$False, it means the edge was indispensable to generating the state, so it has to be True. The process is repeated by selecting randomly another edge, assigning it to False, and checking again if $K$ is satisfiable. The loop continues until all edges are checked and set to False (deleted) or True (kept). We end up with a much-reduced list of edges that, according to $K$, can generate the target state. However, we still need to check if the final solution can generate the state by finding the corresponding weights. The last step of the \textsc{Klaus} algorithm consists of minimizing the infidelity of the resulting graph to find the weights of its edges. 

Many possible solutions may satisfy the $K$ clauses. Moreover, the smallest the graph, the faster the SAT solver, which accelerates the algorithm as it evolves. We can completely trust the logical clauses if they evaluate to False (implying that it is impossible to generate the state with that set of edges). However, the True solutions must still pass the possible interference test between the surviving PMs with the same vertex coloring. It could be the case that a final solution output by \textsc{Klaus} cannot generate the target state because the requiring cancellations cannot occur. This is because all graph PMs constitute a highly coupled system of equations. In some cases, some edges turned out to be indispensable once we minimize the infidelity, so if \textsc{Klaus} has deleted them, then it is not possible to generate the state afterward. In our benchmarks (presented in the next section), we found these cases to be unlikely but they open the path to better understanding the combinatorial nature of this problem and to finding new obstruction clauses to include in our logic instances. We leave the investigation of these constraints for future work.

\section{Benchmarks}\label{sec:benchmarks}

We test and compare the \textsc{Klaus} algorithm with Theseus \cite{krenn2020conceptual}, a purely numerical strategy, to find the minimal graphs that generate a given state. Theseus starts with the fully connected graph and minimizes the infidelity with respect to the target state. In the original proposal, after this minimization, it selects the smallest weight, deletes it (i.e., sets it to zero), and repeats the minimization process until no more weights can be deleted without compromising the infidelity. We found that this approach can be improved significantly by deleting more than one edge at once. In particular, after each minimization, we delete all edges with weights smaller than a certain threshold. 

Although this improved version of Theseus is much faster than the original one, it is not sensitive to those cases where only a subset of weights with similar values can be deleted. Therefore, there is no way to certify that more edges can be removed than trying to delete them one by one, as in the original proposal. Since the goal of these algorithms is to provide a minimal solution, it is necessary to include a final step in Theseus that checks if there is an even smaller solution. 

We try to certify the minimal solution of Theseus following two strategies. Both strategies check if it is possible to remove more edges by proceeding one by one. The first strategy, which we call \emph{Theseus optimization} (TheseusOpt), is performed by following the original Theseus approach, i.e., deleting one edge, minimizing the infidelity, and keeping it if it gets compromised or deleting it definitively otherwise. In the second strategy, called \textsc{Klaus} \emph{optimization} (KlausOpt), we use \textsc{Klaus} instead, i.e., checking if $K$ is still satisfiable when we delete one by one the remaining edges and minimizing the infidelity only at the very end of the algorithm.

We start our benchmarks by checking the performance of these four algorithms (\textsc{Klaus}, Theseus, TheseusOpt, and KlausOpt) with the generation of target states from which we know there exists a graph \cite{gu2019quantumgraphsIII}. We check the computational time that they need and the number of edges of the solution. Since all these algorithms have a heuristic component (the selection of random edges to delete), we run them 25 independent times for each target state to obtain an average performance.

The test states have different entanglement properties quantified by the Schmidt Rank Vector (SRV) \cite{huber2013structure}, a different number of parties $n$, and a different number of basis elements. In particular, we look for the graphs for the GHZ($n$,$d$) states GHZ(4,3) and GHZ(6,2), and states with SRV equal to $(5,4,4)$, $(6,4,4)$, $(6,5,4)$ and $(9,5,5)$. The wave functions of these states are written explicitly in App. \ref{app:states}. The SRV states are composed of three parties. Thus, we will find the graphs of the heralded state, in particular $|\tilde{\psi}\rangle = |\psi\rangle|0\rangle$, where $|\psi\rangle$ is the real target state.

Besides checking if \textsc{Klaus} and Theseus can find states that can be generated from graphs, we also test those states that cannot be exactly constructed this way. These states will be the $GHZ(6,3)$ and two states with SRV equal to $(5,4,4)$ and $(6,4,4)$ different from the above ones. For these states, however, we can obtain approximate solutions by setting those forbidden vertex colorings weights close to zero. Notice that these solutions are forbidden by the logic clauses in \textsc{Klaus}, so we expect that \textsc{Klaus} will have more difficulties finding them.

\begin{figure}[t!]
    \centering
    \includegraphics[width=\linewidth]{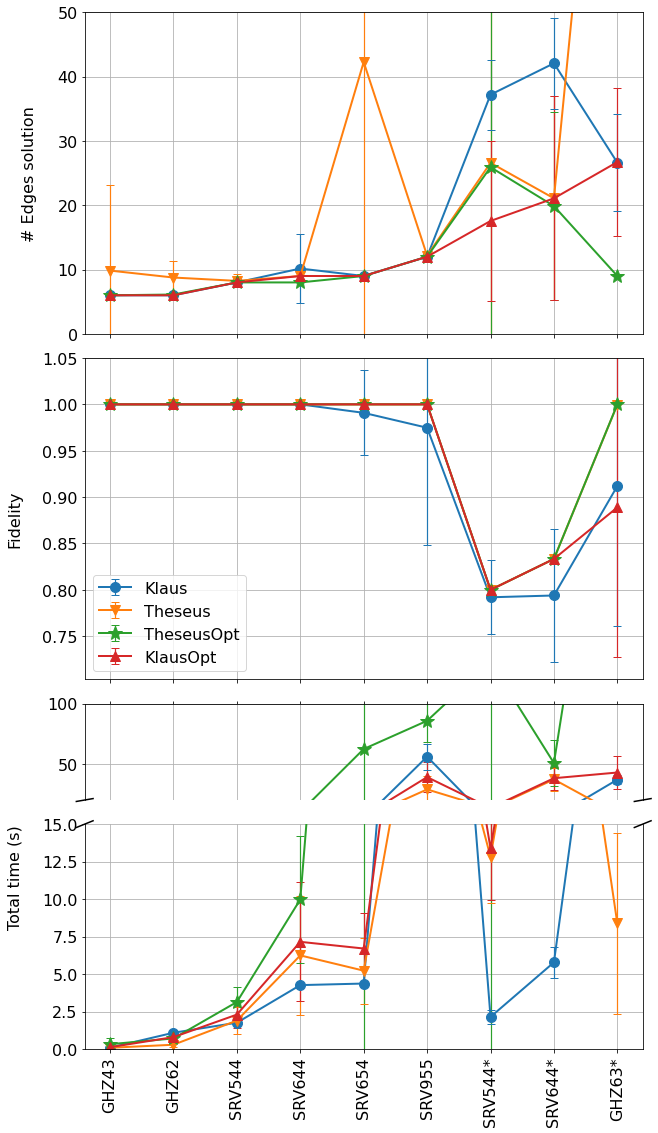}
    \caption{Comparison of the average performance of \textsc{Klaus}, Theseus, TheseusOpt, and KlausOpt algorithms. We take a set of target states that can be generated by graphs and some that cannot (indicated with a * in the plot). We compare the number of edges of the minimal graph solution, the fidelity with respect to the target state, and the total computational time. Since all these algorithms are heuristic, we run each of them 25 times and compute the average and standard deviation of their results. On average, \textsc{Klaus} succeeds in both finding the minimal solution and in spending less computational time on average than the other algorithms. However, it fails to find approximate solutions to those states that cannot be generated by graphs. We expect this result from a propositional logic algorithm, where the clauses $K$ will be False for those approximate solutions. KlausOpt algorithm is significantly better than TheseusOpt, showing the advantage of using a hybrid numerical-logical approach in contrast to purely numeric strategies.}
    \label{fig:benchmark}
\end{figure}

Figure \ref{fig:benchmark} shows the average performance and its standard deviation over 25 independent runs of the four algorithms for the aforementioned target states. The plots show the number of edges of the minimal solution, the fidelity with respect to the target state, and the total computational time (on a 2.4 GHz CPU with 16 GB of RAM). Besides the pure algorithmic optimization time, the computational time for \textsc{Klaus} and KlausOpt includes the generation of the logical clauses. 

We can appreciate how \textsc{Klaus} is, on average, faster than Theseus for those states with no exact graph solution and comparable in general. \textsc{Klaus} finds the minimal solution for those states that can be represented with graphs. However, for those without a graph representation, \textsc{Klaus} obtains solutions with more edges and worse fidelities. We expect this behavior since the logical instances may forbid the aforementioned approximate solutions that Theseus can find. The sometimes big standard deviations are a consequence of the heuristic nature of these algorithms, specially Theseus when it gets trapped in local minima.
In any case, KlausOpt is significantly better than TheseusOpt in terms of the number of edges of the final solution and especially the computational time required, establishing a clear advantage of using the SAT solver instead of multiple numerical minimizations.

\section{Discussion and conclusions}\label{sec:conclusions}

We have shown how logic AI can contribute to the discovery of novel quantum optical setups and experiments. We introduce a Boolean encoding of the graph representation of these setups and present a mapping from the state preparation problem to a $k-$SAT. With this approach, we can check the conjecture that it is not possible to generate a GHZ state of $n$ parties and $d\geq n/2$ dimensions using these experiments. Then, we design a logic-heuristic algorithm, \textsc{Klaus}, which starting from the complete graph, finds the minimal representation that corresponds to the generation of the target state. We benchmark \textsc{Klaus} with the state-of-the-art algorithm Theseus \cite{krenn2020conceptual}, based on numerical optimization. \textsc{Klaus} is on average comparable in execution time or faster than Theseus and it finds the minimal graphs for the different test states. We also show how Theseus, a continuous optimization algorithm, can be improved with the assistance of \textsc{Klaus}, a logic-based algorithm.

At the very end, \textsc{Klaus} has to numerically minimize a loss function consisting of the infidelity between the remaining graph and the target state. However, the process of deleting edges from the fully connected graph simplifies that minimization substantially. There are several potential advantages of using \textsc{Klaus} in contrast to fully-numerical approaches such as the Theseus algorithm: \textit{i)} if $K=$False we know for sure that the graph cannot be exactly generated, while a non-successful purely numerical minimization may imply that we got trapped in local minima; \textit{ii)} the final minimization step involves a small subset of weights, increasing the probability of a successful optimization, in contrast with Theseus, where a minimization involving all weights is performed at the very beginning of algorithm; \textit{iii)} SAT solvers have improved in the last years, becoming a powerful tool in computation that can solve huge problems involving thousands of literals. It makes them a very convenient tool for problems that grow exponentially with the number of parties involved. 

The experimental preparation of quantum states is a key feature in the quantum technologies era. Quantum computing paradigms such as measurement-based quantum computation \cite{raussendorf2003measurement} rely on the initial optimal preparation of highly entangled states. Some quantum machine learning algorithms require the encoding of arbitrary data into the amplitudes of a general quantum state \cite{PhysRevA.94.022342}, including those early proposals that solve a system of linear equations \cite{PhysRevLett.103.150502}. Besides these state preparation applications, the power of the graph representation introduced in \cite{krenn2020conceptual} can also be extended to general quantum operations and quantum circuit design that lead to novel ways to construct, for instance, multilevel multiphotonic experiments \cite{wang2018multidimensional,PhysRevLett.126.230504}, whether integrated \cite{PhysRevLett.123.250503,wang2020integrated} or in bulk optics \cite{PhysRevX.5.041015,PhysRevLett.123.250503,zhong2020quantum}. Although a fully-programmable quantum computer can theoretically prepare any state or perform any unitary operation, not all hardware implementations have direct access to all of the required quantum gates. In this context, providing alternative representations and algorithms based on them will prove valuable in the coming years. Another impactful application of SAT solvers in this context would be the search for limits on success probabilities for quantum state generation or quantum transformations, for instance, those resource states used in quantum computing paradigms such as fusion-based quantum computation \cite{bartolucci2021fusion}. This feat will require handling probabilities (or fractions) suitably as logical clauses.

Although current SAT solvers are extremely efficient and capable of dealing with thousands of literals and clauses, it is worth noting the efforts of quantum and quantum-inspired approaches to solve classical satisfiability problems. In particular, a quantum computing paradigm such as quantum annealing \cite{battaglia2005optimization,su2016quantum} is especially suitable to map classical logical clauses into a quantum Hamiltonian and obtain the solution by adiabatically preparing its ground state. Digital quantum computations can also be programmed to prepare these ground states, even in near-term quantum devices \cite{leporati2007three,farhi2014quantum}. Moreover, quantum-inspired classical techniques such as tensor networks can also be applied to solve SAT problems \cite{garcia2012exact}.

Logic AI, a paradigm proposed in the 50s, is experiencing its expansion in recent years, with the improvements in SAT solvers. Traditionally, it has been mainly used in circuit design, but its applications go beyond that. The increasing interest in understanding concepts such as how a machine learns or how to tackle hard mathematical conjectures has recently promoted this AI subfield. The use of formal reasoning can form fascinating synergies with other approaches. As an example, one can introduce a logic-based piece in a standard ML loss function \cite{xu2018semantic}. Within this work, we present one of the aforementioned synergies by entangling a purely numerical algorithm with a logical one and extend the applicability of logic AI to the design of quantum experiments.

\section*{Code availability}

The Mathematica notebook with \textsc{Klaus} algorithm can be found at \url{https://github.com/AlbaCL/Klaus}.

\section*{Acknowledgements}

A.C.-L. and M.K. acknowledge the insightful discussions with Kevin Mantey, Xuemei Gu, Alex Ravsky and Jakob S. Kottmann.
A.C.-L. and M.K. are thankful for the cozy and friendly environment created (when possible) by Daryl ``Santa" the barman during the hard COVID months in Toronto that contributed substantially to the development of this and other projects.
A.A.-G. acknowledges the generous support from
Google, Inc. in the form of a Google Focused Award.
This work was supported by the U.S. Department of Energy under Award No. DESC0019374 and the U.S. Office of Naval Research (ONS506661). A.A.-G. also acknowledges support from the Canada Industrial Research Chairs Program and the Canada 150 Research Chairs Program. M.K. acknowledges support from the FWF (Austrian Science Fund) via the Erwin Schr\"odinger fellowship No. J4309. 


\bibliographystyle{quantum}
\bibliography{ref}


\onecolumn
\appendix

\section{Boolean algebra and satisfiability} \label{app:boolean}

\begin{figure*}[ht!]
    \centering
    \includegraphics[scale=1]{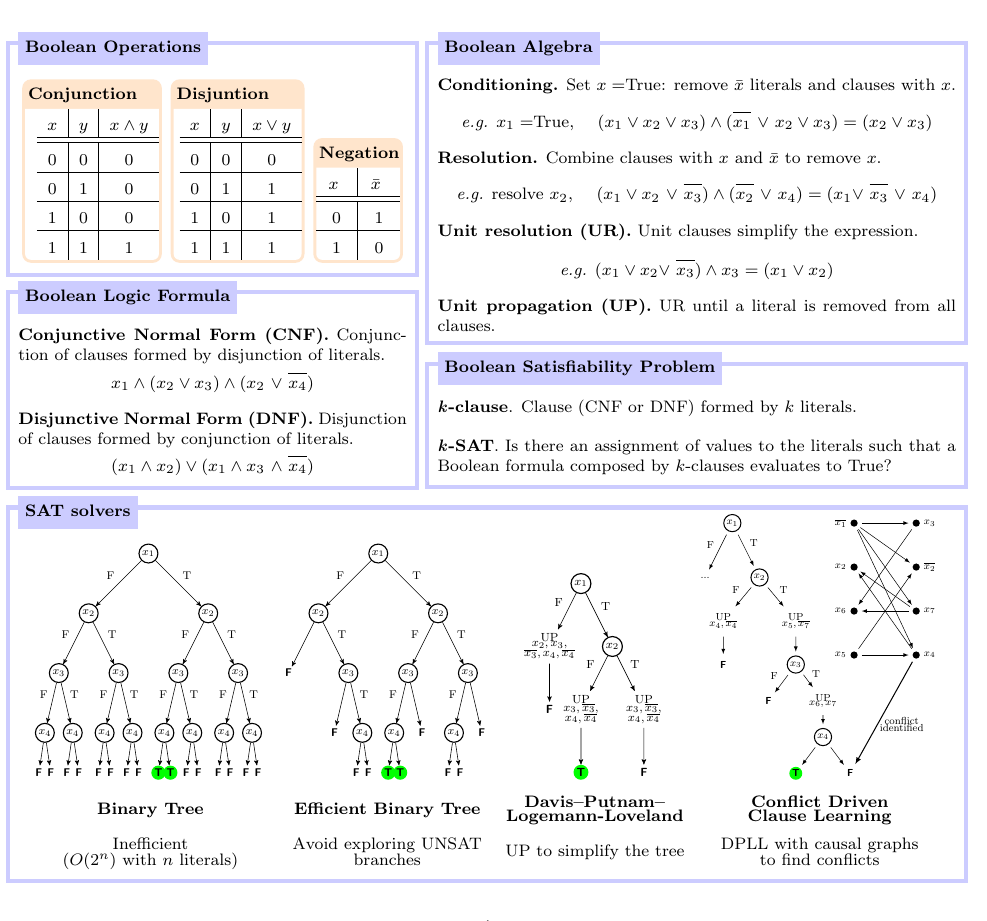}
    \vspace{-0.5cm}
    \caption{\textit{Opening the logic black box.} SAT solvers are extremely sophisticated algorithms capable of dealing with thousands of variables and clauses. They are based on Boolean algebra which variables (called literals) can take two definite values, True-False or 0-1. SAT solvers find the values of these literals that satisfy a Boolean formula (normally written in CNF). If it does not exist a solution, we say the clauses are unsatisfiable (UNSAT). SAT problems are NP-complete, which means it does not exist efficient algorithm that solves them but once the solution is provided, it can be easily verified. However, it is possible to design highly efficient algorithms that go much beyond the naive binary-tree search.}
    \label{fig:SAT}
\end{figure*}

In a Boolean algebra, the variables, called \textit{literals}, can take two Boolean values: True-False or 0-1. 
The available operations on these literals are \textit{disjunctions} $\lor$ (OR), \textit{conjunctions} $\land$ (AND) and \textit{negations} $\bar{x}$ ($\lnot$ or NOT). A \textit{Boolean formula} includes its literals and the operations between them. It is usually more practical to translate a Boolean formula into one of its canonical forms: conjunctive normal form (CNF) or disjunctive normal form (DNF). A CNF expression is a conjunction of \textit{clauses}, each composed of a disjunction of literals. Similarly, a DNF expression is the opposite of a CNF, a disjunction of clauses, where each of them is composed by the conjunction of its literals.

Given a Boolean formula, the satisfiability (SAT) problem consist of finding a literals assignment that satisfies it, i.e. outputs True or 1. The complexity of a SAT problem depends on the structure of its canonical forms, CNF or DNF. This is why the first step toward solving a SAT consists of rewriting the Boolean expression into CNF or DNF. In both cases, we can use logical equivalence rules to find these forms, although this translation can be very costly. In the case of CNF there exist the Tseitin transformation \cite{tseitin1983complexity} which is, in general, more efficient. This is the reason why big SAT problems are normally translated into CNF instead of DNF.

A CNF clause with $k$ literals is called a $k-$clause. A $k-$SAT problem consists of finding if there is any assignment of literals such that a CNF expression composed by $k-$clauses is True. For $k=2$, 2-SAT, the problem is in the P complexity class, meaning it can be solved in polynomial time. For $k>2$, the SAT is an NP-complete problem. This means it can be \textit{verified} in polynomial time, but whether it can 
be solved in such time depends on a solution of a famous complexity theory open problem
whether P=NP. On top of that, any other NP problem can be reduced in polynomial time to $k-$SAT, thus any advances in solving $k-$SAT will impact the whole NP family. As a final remark, although NP is usually used as a synonym of hardness, not all NP problems (or, equivalently, $k-$SAT problems) are hard: the solution of the hardest instance is unknown to be available in polynomial time, but not all problems have them. As a matter of fact, on average, a $k-$SAT problem can be solved relatively quickly and it is actually difficult to find these hard instances to benchmark the SAT solvers. This is why using logic AI and these solvers is a valid strategy even though they are tackling NP problems.

The science of SAT solvers is extremely complex and requires the knowledge and manipulation of logical instances. Once we have our expression in CNF, the SAT solver manipulates it using Boolean algebra trying to find contradictions (such as $x\land\bar{x}$), simplifications, and structures that prevent it to perform a brute-force search. Precisely, a complete SAT solver can use a binary tree approach to check all branches until it finds one that is satisfiable, but it will be highly inefficient. Instead, as it explores the binary tree, it checks if the expression can be simplified and what are the implication relations between the clauses that will trigger a chain reaction depending on the value of one of them. Figure \ref{fig:SAT} shows some SAT solvers examples and some Boolean algebra manipulations that they use. One of the most famous approaches is the Davis–Putnam–Logemann-Loveland (DPLL) algorithm \cite{davis1962machine} and the Conflict Driven Clause Learning (CDCL), from which the algorithms used in this work, the MiniSAT, is based \cite{een2003extensible,MiniSAT}.

\section{From states to graphs} \label{app:graphtostates}

In this appendix, we provide the explicit equivalence between the graph-theoretical representation of photonic experiments and the quantum states generated by them.

The main component of the photonic graph-theoretical representation is the photonic sources, i.e. the SPDC. Each SPDC creates a photon pair at two paths, each with a particular mode. Each SPDC is represented with an edge between two nodes (the paths where the photons are created) and the color of that edge indicates the modes of the photons created.

Mathematically speaking, each SPDC corresponds to the following operation:
\begin{equation}
    \text{SPDC}_{k}(g)=e^{g a_{k}^{\dagger}b_{k}^{\dagger} + g^{*}a_{k}b_{k}}
    \simeq 1 + g a_{k}^{\dagger}b_{k}^{\dagger} + g^{*}a_{k}b_{k} + \cdots,
\end{equation}
where $a$ and $b$ are the labels of the two paths where the photons are created, $k$ is the mode of the photons and $g\ll 1$.

When the setup starts, the first SPDC act on the vacuum state, thus
\begin{equation}
    \left(1 + g a_{k}^{\dagger}b_{k}^{\dagger} + g^{*}a_{k}b_{k}\right)|\varnothing \varnothing\rangle = |\varnothing \varnothing\rangle + g |1_{a_k} 1_{b_k}\rangle,
\end{equation}
i.e. one photon with mode $k$ is created at path $a$ and one photon with mode $k$ is created at path $b$. Imagine that we apply another SPDC on the same paths but with a different mode $l$:
\begin{equation}
    \left(1 + g a_{l}^{\dagger}b_{l}^{\dagger} + g^{*}a_{l}b_{l}\right)\left(|\varnothing \varnothing\rangle + g |1_{a_k} 1_{b_k}\rangle \right)
    = |\varnothing \varnothing\rangle + g \left(|1_{a_k} 1_{b_k}\rangle + |1_{a_l} 1_{b_l}\rangle\right) + \mathcal{O}\left(g^2\right),
\end{equation}
thus we have now a superposition of two photons with mode $k$ and two with mode $l$ at paths $a$ and $b$. If the SPDC creates photons with the same mode as before, $k$, the result will become
\begin{equation}
    \left(1 + g a_{k}^{\dagger}b_{k}^{\dagger} + g^{*}a_{k}b_{k}\right)\left(|\varnothing \varnothing\rangle + g |1_{a_k} 1_{b_k}\rangle\right)
    = (1+g^{*})|\varnothing \varnothing\rangle + 2g |1_{a_k} 1_{b_k}\rangle + g^2|2_{a_k} 2_{b_k}\rangle.
\end{equation}
In any case, we can discard those states containing zero or more than 2 photons since we condition the state on the simultaneous photon detection events in all detectors

Let's see how these manipulations can be analyzed using the graph PMs instead of a four-path photonic example.

\begin{figure}[t!]
    \centering
    \includegraphics[width=0.4\columnwidth]{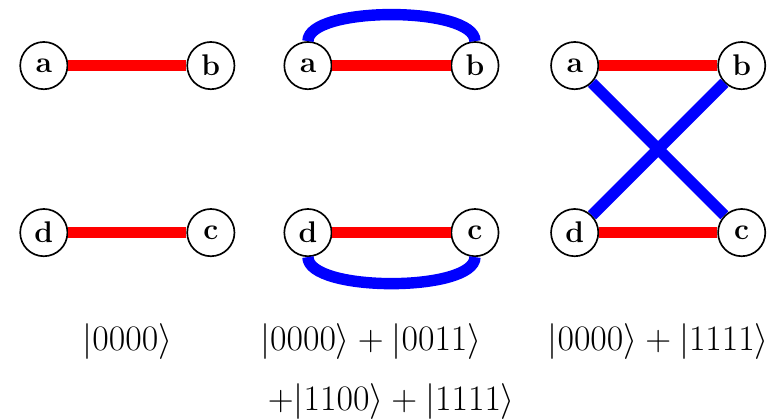}
    \caption{Some examples of four-vertex graphs and the one-photon states that they generate.}
    \label{fig:simpleexample}
\end{figure}

Take the left graph from Fig.\ref{fig:simpleexample}. It corresponds with two SPDC that create photons in the $|0\rangle$ mode in paths $a$, $b$, $c$ and $d$. The state that arrives at the photon detectors is
\begin{eqnarray}
\left(1 + g a^\dagger_0 b^\dagger_0\right)\left(1 + g c^\dagger_0 d^\dagger_0\right)|\varnothing \varnothing\varnothing \varnothing\rangle &=& 
\left(1 + g a^\dagger_0 b^\dagger_0 + g c^\dagger_0 d^\dagger_0 + g^2 a^\dagger_0 b^\dagger_0 c^\dagger_0 d^\dagger_0 \right)|\varnothing \varnothing\varnothing \varnothing\rangle \nonumber\\ &=& 
|\varnothing \varnothing\varnothing \varnothing\rangle + g\left(|00\varnothing\varnothing\rangle + |\varnothing\varnothing00\rangle\right) + g^2|0000\rangle,
\end{eqnarray}
where we avoided the annihilation operators since they are discarded when acting on the vacuum state.
Since we are interested in those photonic states that involve one photon per path, the surviving term is the $|0000\rangle$ state with amplitude $g^2$.

Now, let's take the center graph from Fig.\ref{fig:simpleexample}. After the 0-mode SPDC, we apply 1-mode SPDC on the same pairs of paths. As a result:
\begin{eqnarray}
\left(1 + g a^\dagger_1 b^\dagger_1\right)\left(1 + g c^\dagger_1 d^\dagger_1\right) \left(|\varnothing \varnothing\varnothing \varnothing\rangle + g^2|0000\rangle g\left(|00\varnothing\varnothing\rangle + |\varnothing\varnothing00\rangle\right) \right) \nonumber\\
=|\varnothing \varnothing\varnothing \varnothing\rangle + g\left(|00\varnothing\varnothing\rangle + |\varnothing\varnothing00\rangle\right) + g^2|0000\rangle \nonumber \\
+ g\left(|11\varnothing\varnothing\rangle + |\varnothing\varnothing 11\rangle\right) + g^2|1111\rangle \nonumber \\
+ g^2\left(|0011\rangle + |1100\rangle\right) + \cdots.
\end{eqnarray}
Thus, in those states that contain one photon each, the state generated becomes
\begin{equation}
    g^2\left(|0000\rangle + |0011\rangle + |1100\rangle + |1111\rangle\right).
\end{equation}

Finally, let's consider the right graph from Fig.\ref{fig:simpleexample}. The second row of SPDC is applied on different pairs of paths than the first one. Thus:
\begin{eqnarray}
\left(1 + g a^\dagger_1 c^\dagger_1\right)\left(1 + g b^\dagger_1 d^\dagger_1\right)\left(|\varnothing \varnothing\varnothing \varnothing\rangle + g^2|0000\rangle + g\left(|00\varnothing\varnothing\rangle + |\varnothing\varnothing00\rangle\right) \right) \nonumber \\
=|\varnothing \varnothing\varnothing \varnothing\rangle + g\left(|00\varnothing\varnothing\rangle + |\varnothing\varnothing00\rangle\right) + g^2|0000\rangle \nonumber \\
+ g\left(|1\varnothing1\varnothing\rangle + |\varnothing 1\varnothing 1\rangle\right) + g^2|1111\rangle + \cdots.
\end{eqnarray}
This time, the surviving terms are two. $g^2\left(|0000\rangle + |1111\rangle\right)$. 

So, we can draw three conclusions from this analysis:
\begin{enumerate}
    \item Each graph PM generates a basis element corresponding to the modes (colors) incident to the paths (vertices).
    \item The final superposition state corresponds to adding all graph PMs.
    \item The edge weight corresponds to some power of the SPDC coupling $g$. The exponent corresponds to $n/2$ where $n$ are the total paths.
\end{enumerate}

\section{Formal definitions of graph representation of optical experiments} \label{app:formalgraph}

Let us formulate the graph-based representation of optical experiments in a more formal way (for a more detailed mathematical description, check Ref.\cite{krenn2019questions}).

Given a graph with $n$ vertices and a set of undirected edges $E$, a perfect matching (PM, in plural PMs) corresponds to a set of edges $e\in E$ such that each vertex is matched with exactly one edge. For weighted graphs, i.e., for graphs where each edge has an associated weight $w\in\mathbb{C}$, the total weight of a PM corresponds to the product of the weights that forms it. We can add more degrees of freedom by associating another property to the edges: color. We assume that each edge of $G$ contains up to two colors (\emph{bi-chromatic} graphs). A bi-chromatic edge with a color pair $(\alpha,\beta)$ will join two vertices $(i,j)$, giving color $\alpha$ to vertex $i$ and color $\beta$ to vertex $j$. Then, each edge contains five properties: the two vertices it joins, the corresponding colors that deliver to each vertex, and its complex weight. 

We label each edge with $e_{ij}^{\alpha\beta}$, where $(i,j)$ with $i<j$ are the vertex pair and $(\alpha,\beta)$ are the corresponding colors. Similarly, the weights of each edge will be labelled as $w_{ij}^{\alpha\beta}\in \mathbb{C}$. Thus, a PM $P$ and its associated weight $w_{P}$ are defined as
\begin{eqnarray}
    P(c)&=&\{e_{ij}^{\alpha\beta}\} \ \text{for} \ i,j\in V \ i\neq j, \\
    w_{P}(c)&:=&\prod_{(i,j)\in V} w_{ij}^{\alpha\beta},
\end{eqnarray}
where $c$ is the color combination inherited by the vertices. For an example, take the first graph from \ref{fig:obstruction}a. The $n=6$ graph with edges $E= \{e_{ab}^{00}, e_{cd}^{00}, e_{ef}^{11}\}$ form one PM $P$ with weight $w_{P}(c)=w_{ab}^{00} w_{cd}^{00} w_{ef}^{11}$ and $c=(0,0,0,0,1,1)$.

A general graph may contain several PMs. In particular, a complete graph with $n$ vertices contains $(2n-1)!!=(2n-1)!/(2^{n-1}(n-1)!)$ PMs. If each edge of the graph has the extra color degrees of freedom, the number of PMs increases to $d^{n}(2n-1)!!$, where $d$ is the number of different colors. Therefore, there could be more than one PM with the same inherited vertex coloring, i.e. the same color combination inherited by the vertices from the bi-colored edges that touch them. As explained before, each color vertex combination corresponds to the generation of a basis state. Thus, to obtain the total basis state amplitude, we need to sum up the weights of all PMs that generate it. The \emph{weight of a vertex coloring} $c$ of a graph is
\begin{equation}
    W(c):=\sum_{P\in\mathcal{M}}\prod_{p\in P}w_{p}(c),
\end{equation}
where $\mathcal{M}$ is the set of perfect matching of G with the same coloring $c$ and $w_{p}$ are the corresponding PM weights of each  $P\in\mathcal{M}$. Coming back to the previous example, if we add to the list of edges the edges $e_{ac}^{00}$ and $e_{bd}^{00}$, the resultant graph contains $E= \{e_{ab}^{00}, e_{cd}^{00}, e_{ef}^{11}, e_{ac}^{00}, e_{bd}^{00}\}$ and thus it generates a second PM, the second one shown in Fig.\ref{fig:obstruction}a. That PM has the same vertex coloring as the previous one, $c=(0,0,0,0,1,1)$. Thus, the weight of that vertex coloring is $W(c)=w_{P}(c)=w_{ab}^{00} w_{cd}^{00} w_{ef}^{11} + w_{ac}^{00} w_{bd}^{00} w_{ef}^{11}$.

\section{Logic clauses construction}\label{app:formallogic}

For a given PM $P$, its Boolean expression becomes
\begin{equation}
    b_{P}(c)=\bigwedge_{\{i,j\}\in P} e_{ij}^{\alpha\beta},
\end{equation}
where $e_{ij}^{\alpha\beta}$ are the graph edges (and the Boolean literals), $i<j$ are the graph vertices and $c$ coloring will be defined by the particular colors $\alpha$ and $\beta$ associated with each edge in canonical order.

We require that at least one PM exist for each vertex coloring that appears in the target state. Thus, the collection of clauses that encode this logical statement becomes
\begin{equation}
    B(c) = \bigvee_{P\in\mathcal{M}}\bigwedge_{\{i,j\}\in P} e_{ij}^{\alpha\beta},
\end{equation}
where $\mathcal{M}$ is the set of PMs with $c$ vertex coloring and $B_{c}$ is False only if all PMs are False, and True otherwise. As required, we need at least one PM with vertex coloring $c$ to generate the state with that coloring. In total, we need that this property is fulfilled for each of the vertex colorings that appear in the target state that we want to generate. Thus, the total logical clause for the target state elements becomes
\begin{equation}
    S = \bigwedge_{c\in \mathcal{C}} B(c) = \bigwedge_{c\in \mathcal{C}} \bigvee_{P\in\mathcal{M}_{c}}\bigwedge_{\{i,j\}\in P} e_{ij}^{\alpha\beta},
    \label{eq:state_clause}
\end{equation}
where $\mathcal{C}$ is the set of vertex coloring that appear in the target state and $\mathcal{M}_{c}$ are the set of PMs for each of these colorings.

To encode the obstructions to those basis elements that do not appear in the target state, we use the following logic: if all PMs except one that generates those basis elements are False, the remaining one has to be False as well. However, other possibilities, e.g. two or more are True, are allowed since there can be cancellations between the weights of these PMs.
For each of these forbidden basis elements, we encode this logical statement in the following way:
\begin{equation}
    C(c)= \bigwedge_{k\in\mathcal{M}} \overline{b_{k}(c)}\lor \bigwedge_{\substack{P\in\mathcal{M}\\ P\neq k}} b_P(c),
    \label{eq:obstr}
\end{equation}
where $\mathcal{M}$ are the set of PMs with vertex coloring $c$. Take a subset of all PMs with the same vertex coloring consisting of all PMs except one. If all PMs of this subset is False, its conjunction will be False. Therefore, to $C(c)=$True, the remaining PM must evaluate to False as well. For example, imagine we have three PMs with a vertex coloring $c$ that must not appear in the target state, namely $PM_{1}$, $PM_{2}$ and $PM_{3}$. If $PM_{2}=PM_{3}=$False, then $PM_{2}\lor PM_{3}=$False. As a consequence, $PM_{1}=$False, so $\overline{PM_{1}}=$True in order to obtain $C(c)=$True. 

Considering all basis elements that do not appear in the target state, the \emph{obstruction} clause becomes
\begin{equation}
    C= \left(\bigwedge_{o\in\mathcal{O}}C(o)\right),
    \label{eq:final_obst}
\end{equation}
where $\mathcal{O}$ is the set of vertex colorings that do not appear in the state. This clause evaluates to True only when all subclauses are fulfilled, i.e. each $C(c)=$True.

\section{Benchmark states} \label{app:states}

\begin{figure*}
     \subfloat[\label{subfig2:a}]{%
         \centering
         \includegraphics[width=0.3\linewidth]{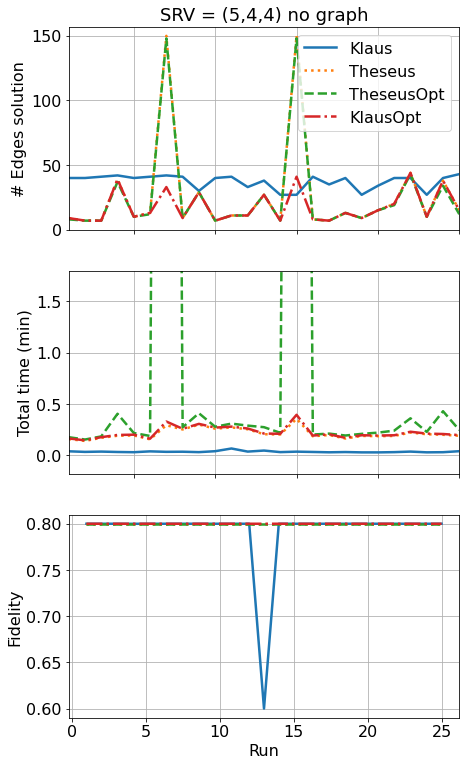}
      }
      \hfill  
    \subfloat[\label{subfig2:b}]{%
         \centering
         \includegraphics[width=0.3\linewidth]{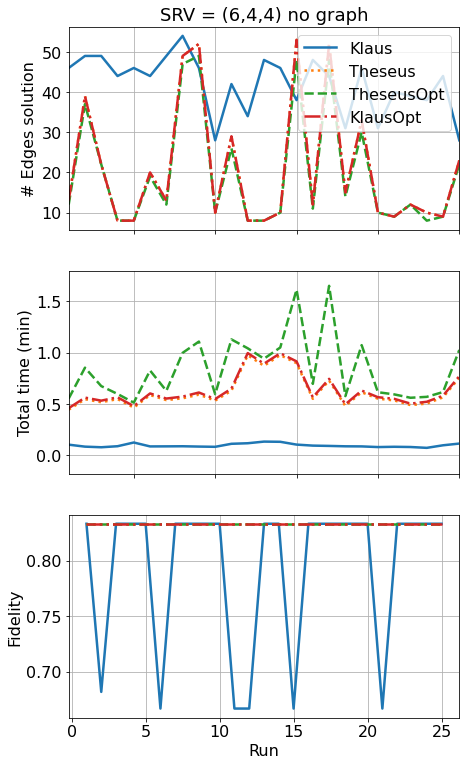}
     }
     \hfill
     \subfloat[\label{subfig2:c}]{%
         \centering
         \includegraphics[width=0.3\linewidth]{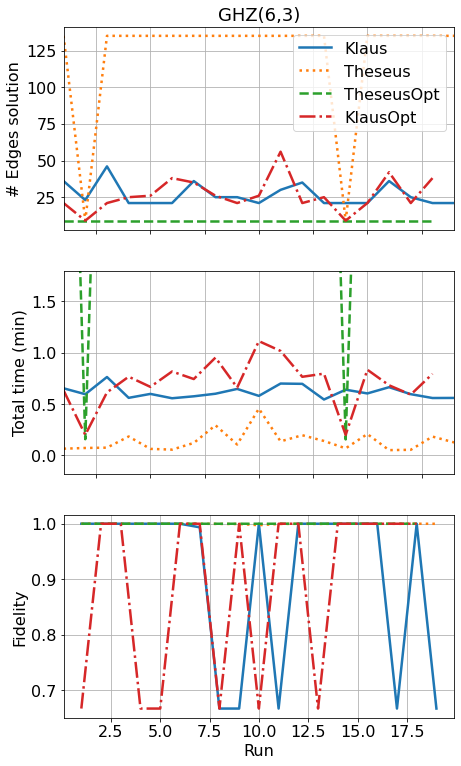}
     }
        \caption{Explicit data states from which there does not exist a graph}
        \label{fig:data_states_no_graph}
\end{figure*}

The states used in the benchmarks are the GHZ states
\begin{equation}
    |GHZ_{n,d}\rangle=\frac{1}{\sqrt{d}}\sum_{k=0}^{d-1}|k\rangle^{\otimes n},
\end{equation}
in particular the GHZ states for $n=4,6,8$ and local dimension 2 or 3:
\begin{eqnarray}
|GHZ_{4,3}\rangle &=& \frac{1}{\sqrt{3}}\left(|0000\rangle + |1111\rangle + |2222\rangle\right), \\
|GHZ_{6,2}\rangle &=& \frac{1}{\sqrt{2}}\left(|000000\rangle + |111111\rangle \right), \\
|GHZ_{8,2}\rangle &=& \frac{1}{\sqrt{2}}\left(|00000000\rangle + |11111111\rangle\right).
\end{eqnarray}

Besides these states, we also consider highly entangled states with different Schmidt Rank Vector (SRV) \cite{huber2013structure}. This figure of merit is well-defined for states consisting of 3 parties. That is why to generate these states we introduce a fourth party in the $|0\rangle$ state, i.e. we look for the graph that generates the state $|\psi\rangle|0\rangle$, where $|\psi\rangle$ is the true target state. The explicit wave functions of these states are
\begin{eqnarray}
|\Psi_{544}\rangle &=& \frac{1}{\sqrt{5}}\left(|000\rangle + |111\rangle + |222\rangle + |330\rangle + |413\rangle\right),
\\
|\Psi_{644}\rangle &=& \frac{1}{\sqrt{6}}\left(|000\rangle + |111\rangle + |222\rangle +|330\rangle + |413\rangle +|512\rangle  \right),
\\
|\Psi_{654}\rangle &=& \frac{1}{\sqrt{6}}\left(|000\rangle + |111\rangle + |222\rangle + |330\rangle +|440\rangle + |513\rangle\right), 
\\
 |\Psi_{955}\rangle &=& \frac{1}{\sqrt{9}}\left(|000\rangle + |111\rangle + |222\rangle + |303\rangle + |404\rangle + |505\rangle + |631\rangle + |741\rangle + |841\rangle \right). 
\end{eqnarray}
The subindex of the $|\psi\rangle$ states indicates their SRV.

We also use two states with SRV$=(5,4,4),(6,4,4)$ for which it does not exist exact graph that generates them. These states are
\begin{eqnarray}
|\Psi_{544}^{*}\rangle &=& \frac{1}{\sqrt{5}}\left(|000\rangle + |111\rangle + |222\rangle + |333\rangle + |401\rangle\right),
\\
|\Psi_{644}^{*}\rangle &=& \frac{1}{\sqrt{6}}\left(|000\rangle + |111\rangle + |222\rangle +|310\rangle + |420\rangle +|533\rangle  \right).
\end{eqnarray}
We also try to generate the GHZ(6,3), which we know is not representable by an exact graph,
\begin{equation}
    |GHZ_{6,3}\rangle = \frac{1}{\sqrt{3}}\left(|000000\rangle + |111111\rangle + |222222\rangle \right).
\end{equation}

\begin{figure*}
     \centering
     \subfloat[\label{subfig:a}]{%
         \centering
         \includegraphics[width=0.3\linewidth]{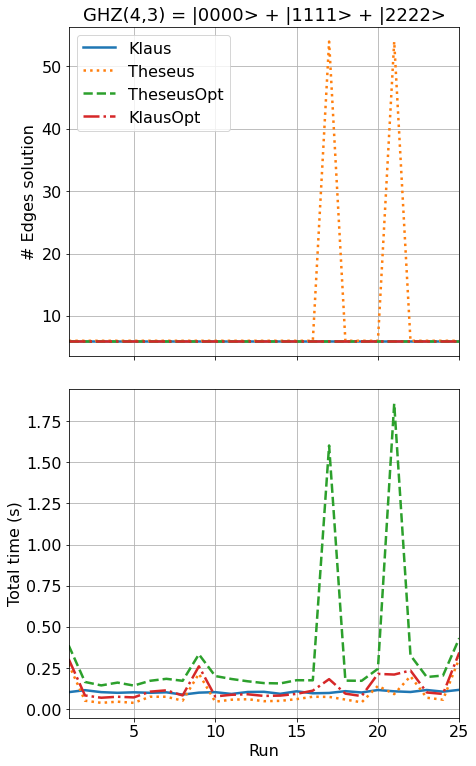}
     }
     \hfill
     \subfloat[\label{subfig:b}]{%
         \centering
         \includegraphics[width=0.3\linewidth]{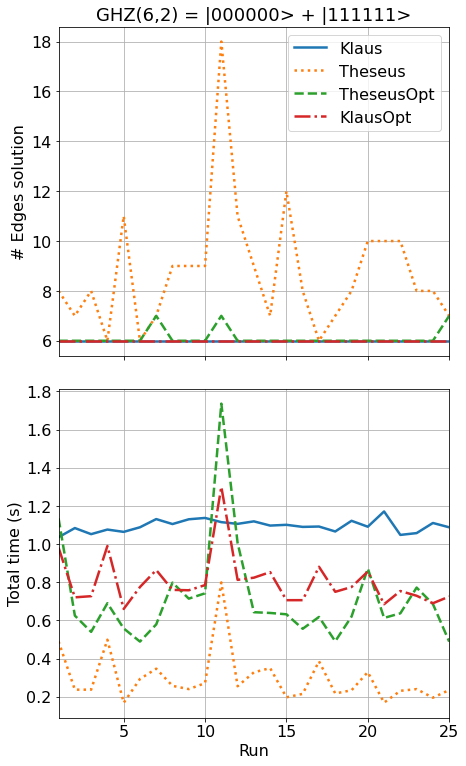}
    }
    \hfill
    \subfloat[\label{subfig:c}]{%
         \centering
         \includegraphics[width=0.3\linewidth]{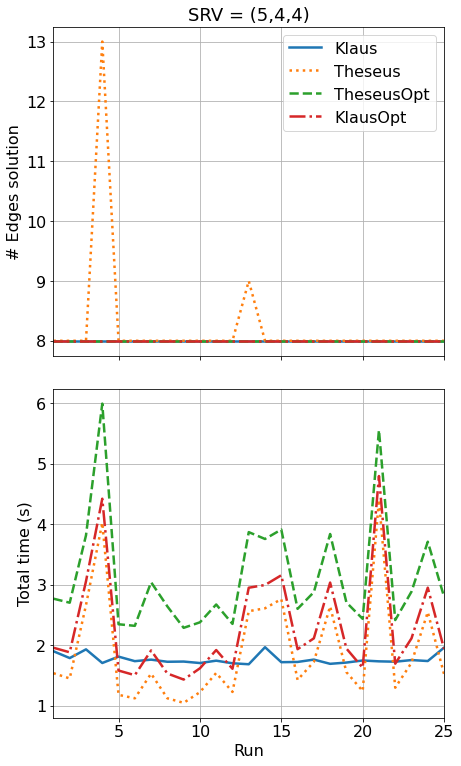}
     }\\
     \vspace{1cm}
     \subfloat[\label{subfig:d}]{%
         \centering
         \includegraphics[width=0.3\linewidth]{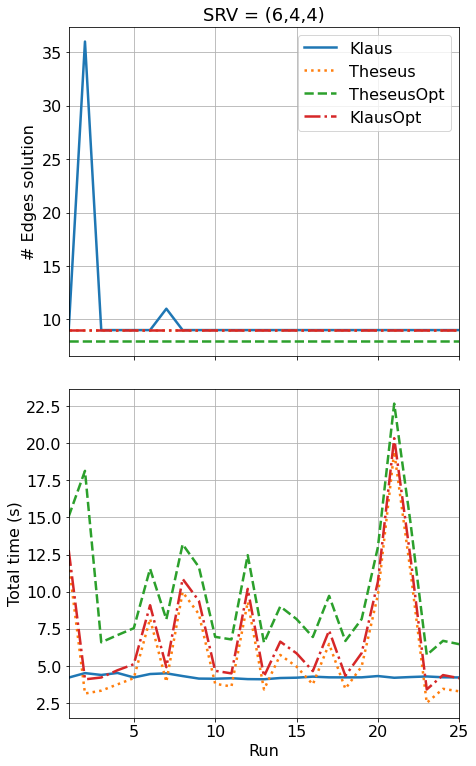}
      }
      \hfill  
    \subfloat[\label{subfig:e}]{%
         \centering
         \includegraphics[width=0.3\linewidth]{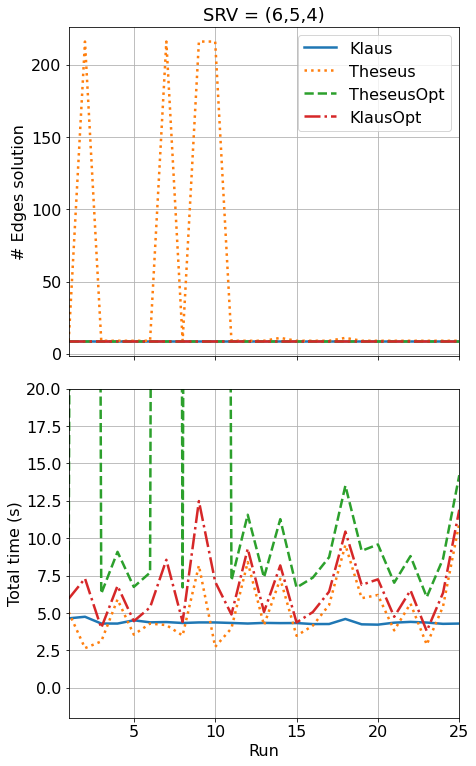}
     }
     \hfill
     \subfloat[\label{subfig:f}]{%
         \centering
         \includegraphics[width=0.3\linewidth]{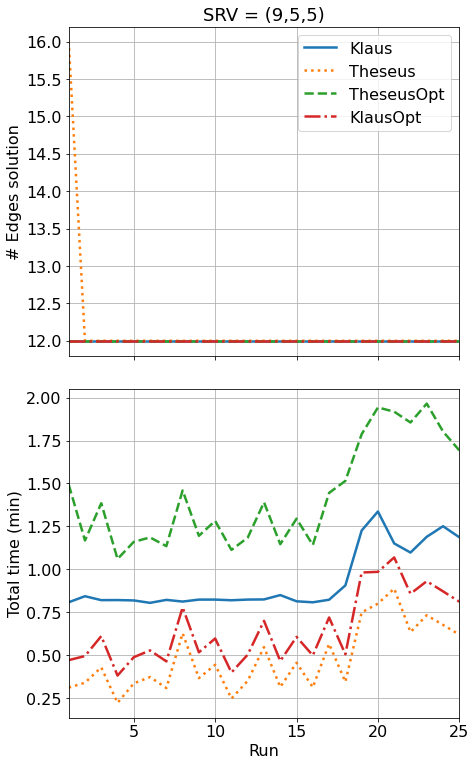}
     }
        \caption{Explicit data states from which there exists a graph}
        \label{fig:data_states_graph}
\end{figure*}

We run the algorithms benchmarks 25 times for each target state and present the average performance in the main article. Figures \ref{fig:data_states_no_graph} and \ref{fig:data_states_graph} show the results of each of these runs.%

\end{document}